\documentstyle[12pt]{article}
\input{psfig}
\begin{document}
\newcommand {\ber} {\begin{eqnarray*}}
\newcommand {\eer} {\end{eqnarray*}}
\newcommand {\bea} {\begin{eqnarray}}
\newcommand {\eea} {\end{eqnarray}}
\newcommand {\beq} {\begin{equation}}
\newcommand {\eeq} {\end{equation}}
\newcommand {\state} [1] {\mid \! \! {#1} \rangleg}
\newcommand {\sym} {$SY\! M_2\ $}
\newcommand {\eqref} [1] {(\ref {#1})}
\newcommand{\preprint}[1]{\begin{table}[t] 
           \begin{flushright}               
           \begin{large}{#1}\end{large}     
           \end{flushright}                 
           \end{table}}                     
\def\Acknowledgements{\bigskip  \bigskip {\begin{center} \begin{large}
             \bf ACKNOWLEDGMENTS \end{large}\end{center}}}

\newcommand{\half} {{1\over {\sqrt2}}}
\newcommand{\dx} {\partial _1}

\def\Dslash{\not{\hbox{\kern-4pt $D$}}}
\def\cmp#1{{\it Comm. Math. Phys.} {\bf #1}}\begin{titlepage}
\titlepage
\vskip 1cm
\centerline{{\Large \bf A new "polarized version" of the Casimir }}
\centerline{{\Large \bf Effect is measurable}}
\vskip 1cm

\centerline{O. Kenneth and  S. Nussinov}

\vskip 1cm
\begin{center}
\em School of Physics and Astronomy
\\Beverly and Raymond Sackler Faculty of Exact Sciences
\\Tel Aviv University, Ramat Aviv, 69978, Israel

\end{center}
\vskip 1cm
\begin{abstract}
We argue that the exactly computable, angle dependent, Casimir force between parallel plates with different directions of conductivity can be measured.

\end{abstract}
\end{titlepage}

Some time ago we presented [N] heuristic arguments motivated by the
"vacuum pressure" picture[Mi] for the Casimir force[C]. These arguments
suggested that the attraction between parallel plates conducting along
different directions, decreases monotonically as a function of
$\gamma$ the angle between these directions. This angle dependent
effect was then evaluated exactly[KN1,KT] as part of the thesis work
of O.Kenneth. The (euclidean) path integral technique\footnote
{These techniques were introduced some time ago and extensively used by
 Kardar and collaborators (K1)(K2).  The specific application for the
 new angle dependent effect (KN1) to the effect for general planar geometry(K),
 and to motivate the attraction between disjoint objects of similar       
 ${\epsilon/\mu}$  ratios (KN2)are however new .} used for this purpose was usefull for computing and understanding a large variety of Casimir related issues[KT,KN2].
Our main purpose here is to note that the angle dependent Casimir effect (which is present in the electromagnetic case because of the polarization degree of freedom) is not only exactly computable. Rather we argue that it is also measurable with a precision of few percent-the precision of recent experiments[L,M] of the ordinary Casimir effect between (isotropically) conducting parallel plates. 

Let us briefly recall the derivation of the angle dependent Casimir
effect. The euclidean\footnote{The Casimir problem for two
 conducting surfaces  $\Sigma_1$ and $\Sigma_2$ is fully                 defined by space like vectors
 only:the vectors $x_i$ and $y_j$  connecting points on the two conductors               
  to their respective centroids, and $a$ the relative displacement of
  the latter.
  Hence there can be no obstruction to complete Wick rotation and
 formulation of the problem via euclidean action and partition function.}
 partition function in the presence of two disjoint  conducting surfaces $\Sigma_1$ and $\Sigma_2$ can be written as:$$
\int{\cal D}J_{(1)}^\mu(x){\cal D}J_{(2)}^\mu(y)\exp\left\{ -\int J^\mu_{(1)}(x){dxdx'\over(x-x')^2}J^\mu_{(1)}(x')
 -2\int J^\mu_{(1)}(x){dxdy\over(x-y)^2}J^\mu_{(2)}(y)\right.$$\beq\label{jjj}\left.
-\int J^\mu_{(2)}(y){dydy'\over(y-y')^2}J^\mu_{(2)}(y')
\right\}\eeq
\vspace{5mm}
In the above $x,x'\in\tilde{\Sigma}_1$ and $y,y'\in\tilde{\Sigma}_2$ with $\tilde{\Sigma}_i=\Sigma_i\bigotimes${time axis}. This expression is obtained by starting with the Maxwell (euclidean) action $\int F_{\mu\nu}^2$, using
$$\int_{x\in\tilde{\Sigma_1}}A_\mu(x)J_{(1)}^\mu d\sigma_1+\int_{x\in\tilde{\Sigma_2}}A_\mu(x)J_{(2)}^\mu d\sigma_2$$
with the (conserved) currents $J_{(1)}^\mu,J_{(2)}^\mu$ serving as lagrange multipliers forcing the boundary conditions $E_\|=0$ on $\Sigma_1,\Sigma_2$ respectively, doing the gaussian ${\cal D}A_\mu$ integration, and utilizing $\partial_\mu J^\mu=0$ to justify the choice $\Delta^{\mu\nu}(x-y)={g^{\mu\nu}\over(x-y)^2}$ (see ref[KT,KN1] for details).
The $(x-y)^{-2}$ coefficients in the quadratic form in the currents in eq(\ref{jjj}) above is clearly the source of divergences in Casimir energy evaluation.
These divergences arise however {\bf only} from products of two $J^{(1)}$`s at near by points on $\Sigma_1$ or of two $J^{(2)}$`s at near by points on $\Sigma_2$ . Let us divide out ${\cal Z}(a)$ by ${\cal Z}(a\rightarrow\infty)$ which from the definition of ${\cal Z}=e^{-E(a)T}$
with $T$ the size of the time interval is equivalent to subtracting from the total Casimir energy of the system the separate energies of the two conductors $\Sigma_1$ and $\Sigma_2$.
The infinite contributions $\lim_{x_i\rightarrow x_j}J_i^{(1)}J_j^{(1)}\over(x_i-x_j)^2$ clearly divide out for each point $x_i\in\Sigma_1$ etc.
Hence no divergences are expected\footnote{ This is in clear contrast to the casimir energy of each of
the two conductors separately.         
   The latter diverge and
 require careful regularization.  For the mixed products or the mutual
 casimir energy $a$ serves as
 a regulator.  While we are appealing to the lore of               
 renormalization theory and identify all the divergent parts with      the     energies of the separte conductors, a rigorous
             proof that the $Z(a)/Z(\infty)$ ratio (or the $E(a)- E ( \infty)$  is independent of the
 scheme of renormalization will not be supplied here.}

and
\beq\label{uu}{e^{-E(a)T}\over e^{-E(\infty)T}}={\int\prod dJ^{(1)}(x)\prod dJ^{(2)}(y)e^{-\int{J^{(1)}(x)J^{(1)}(x')\over(x-x')^2}}e^{-\int{J^{(1)}(x)J^{(2)}(y)\over(x-y)^2}}
e^{-\int{J^{(2)}(y)J^{(2)}(y')\over(y-y')^2}}\over{\int\prod dJ^{(1)}(x)\prod dJ^{(2)}(y)e^{-\int{J^{(1)}(x)J^{(1)}(x')\over(x-x')^2}}e^{-\int{J^{(2)}(y)J^{(2)}(y')\over(y-y')^2}}}}\eeq
 is finite and well defined (the coefficient of the mixed $J^{(1)}(x)J^{(2)}(y)$ product $1\over (x-y)^2$ is bound by $1\over a^2$, with $a=\min\{ |x-y| ; x\in\Sigma_1,y\in\Sigma_2\}$ the minimal distance between the conductors and is finite)
The expression in eq(\ref{uu}) can serve as a meaningful starting point for numerical estimates or general considerations[KN2]. The latter strongly suggest that $E(a)$ is monotonic and hence that the Casimir force between two disjoint conductors is attractive at all distances.
For the special case at hand $\Sigma_1$ and $\Sigma_2$ are two infinite plates parallel to the $x,y$ plane and separated by $|\vec{a}|=|a\hat{z}|=a$. The general expression of eq (\ref{uu}) becomes:

  \beq \label{actw} {\cal Z}(a)=\int{\cal
  D}J\exp\left(-{1\over{8\pi^2}}\int d^3xd^3y \left({{\vec J_1(x)\cdot \vec
    J_1(y)+\vec J_2(x)\cdot \vec J_2(y)}\over{(x-y)^2}}+ {{2\vec
    J_1(x)\cdot \vec J_2(y)}\over{(x-y)^2+a^2}}\right)\right)\eeq 
     With $\vec J_{1,2}$  ordinary 3-vectors in the
3-dimensional euclidean space $\vec{x}=(x,y,t)$. Fourier transforming in $\vec{x}$ we obtain:
\beq \label{actf} {\cal Z}=\int{\cal D}J\exp\left(-{1\over{8\pi^2}}4\pi^2\int d^3k \left({{\vec
      J_1(\vec k)\cdot \vec J_1(-\vec k)+\vec J_2(\vec k)\cdot \vec J_2(-\vec k)}\over k}+{{2\vec J_1(\vec k)\cdot \vec J_2(-\vec k)}\over k}e^{-ka}\right)\right)\eeq where $\vec k=(k_x,k_y,k_t),k=|\vec k|$ and we used translation invariance and $\int d^3xe^{-ikx}{1\over x^2+a^2}=\pi{e^{-ka}\over k}$.
 In the usual case of two conducting parallel plates both $\vec J_1(k)$ and $\vec J_2(k)$
 have two degrees of freedom corresponding to the two transverse directions which
 satisfy current conservation condition: $\vec k\cdot\vec J=0$. In the case of specific conduction directions  $\vec J_1(k)$ and likewise $\vec J_2(k)$ have only one allowed nonzero component determined by
 current conservation and by the demand that its spatial part
 $(J_x,J_y)$ is along the direction of conduction. Let us denote the
 cosine of the angle between the directions of $\vec J_1(\vec
 k)$ and $\vec J_2(\vec k)$ by $\alpha (\hat{k})$. 
 $\vec J_1,\vec J_2$ are vectors in the 3-dimensional Euclidean
 space $(k_x,k_y,k_t)$. Using ordinary geometry we find an explicit
 expression for $\alpha (k)$ in terms of the direction of $\vec k$ and
 of the conduction directions in the two plates
 \beq\label{alp}
 \alpha^2={{[\cos\gamma-\sin^2\theta\cos\varphi\cos(\varphi-\gamma)]^2}
\over{(1-\sin^2\theta\cos^2\varphi)(1-\sin^2\theta\cos^2(\varphi-\gamma))}}\eeq
where $\vec{k}=(k_x,k_y,k_t)=(k\sin\theta\cos\varphi,k\sin\theta\sin\varphi,k\cos\theta)$ and $\gamma$ is the angle between the directions of conductance in the two plates

Then we can write 
\beq \label{sc} {\cal Z}=\int{\cal D}J\exp\left(-{1\over 2}\int d^3k \left({{
      J_1(k)J_1(-k)+J_2(k)J_2(-k)}\over k}+{{2\alpha(k)J_1(k)J_2(-k)}\over k}e^{-ka}\right)\right)
\eeq
where the $J$'s appearing in the last equations are scalars. $J_1(k)^*=J_1(-k)$ and $J_2(k)^*=J_2(-k)$ forms the reality condition on J. Since the
action is quadratic, $\cal Z$ is given by the corresponding determinant
which is just the product of the two-dimensional determinants
corresponding to the various value of $\vec k$. Hence
\beq {{\cal Z}(a)\over{\cal Z}(\infty)}={\prod}_k \det{\left (\begin{array}{cc} {1\over
 k}&{{\alpha(\hat{k})\over k}e^{-ka}}\\{\alpha(\hat{k})\over
      k}e^{-ka}&{1\over k}\end{array}\right )}^{1\over 2}/
{\prod}_k \det{\left (\begin{array}{cc} {1\over
 k}&0\\ 0&{1\over k}\end{array}\right )}^{1\over 2}\eeq

and

\bea\label{det} \ln { {\cal Z}(a)\over {\cal Z}(\infty)}&=&{1\over 2}\ln\det\left(\cdots\right)-{1\over 2}\ln\det\left(\cdots\right)\nonumber\\&=&{1\over 2}AT\int{{ d^3k}\over{(2\pi)^3}}\ln\left\{ \left |\begin{array}{cc} {1\over
 k}&{{\alpha(\hat{k})\over k}e^{-ka}}\\{\alpha(\hat{k})\over
      k}e^{-ka}&{1\over k}\end{array}\right |-\ln{1\over k^2}\right\} \nonumber
\\&=&{1\over 2}
AT\int {{d^3k}\over{(2\pi)^3}}\ln(1-{\alpha
   (\hat{k})}^2e^{-2ka})
\eea

 where the area-time $AT$ came from density of states factor. It corresponds to having ${\sum}_k\rightarrow V\int{d^3k\over (2\pi)^3}$ the usual quantization of continuous modes in a box of volume V. Also note that a factor $1\over 2$ survives since we integrate over $dRe(J_1(k)),dIm(J_1(k))$ but only over half of the $\vec k$ values say with $k_x>0$. As expected the last integral is well
 defined and convergent. 

Identifying $\ln {\cal Z}=ET$ we get finally:  \beq
 \label{fr} E/A={1\over 2}\int{{d^3k}\over{(2\pi)^3}}\ln(1-{\alpha
   (\hat{k})}^2e^{-2ka})\eeq  
 using integration by parts this can (see ref[KN1] for more details) be written as
\beq \label{fr2} E/A=
-{1\over{48a^3}}\int{{d^3k}\over{(2\pi)^3}}{k{\alpha(\hat{k})}^2\over{e^k-{\alpha
   (\hat{k})}^2}}\eeq

Using (\ref{alp}) our final result is\footnote{ It is possible  to derive the angle dependent effect directly via mode
 summation analogous to the original casimir approach (K.T).}:
\bea\label{rtg} E/A&=&-{1\over 48a^3}{1\over (2\pi)^3}\int_0^{\infty} dk\int_0^{2\pi}
d\varphi\int_0^\pi d\theta\times\\
&\times&\left\{k^3\sin\theta{\left(\cos\gamma-\sin^2\theta\cos\varphi\cos(\varphi-\gamma)\right)^2}
\over{\left(1-\sin^2\theta\cos^2\varphi\right)\left(1-\sin^2\theta\cos^2(\varphi-\gamma)\right)}
{e^k-\left(\cos\gamma-\sin^2\theta\cos\varphi\cos(\varphi-\gamma)\right)^2}\right\}\nonumber \eea
 
For $\gamma=0$ we have $\alpha\equiv 1$ and 
 $E/A=-{1\over 48a^3}\int{4\pi\over (2\pi)^3}{k^3dk\over e^k-1}=-{1\over 96 a^3\pi^2}{\sum}_n\int_0^{\infty}k^3e^{-nk}dk=-{6\sum{1\over n^4}\over 96a^3\pi^2}={-{6{\pi}^4\over 90}\over 96a^3{\pi}^2}=-{\pi^2\over 1440a^3}$
The last expression is exactly half the 

ordinary Casimir energy for isotropic conductivity. 
 The extra factor  of two is expected since $J_1$ and $J_2$ ( and the
 field they cause to vanish on $\Sigma_1$ and $\Sigma_2$) 
are scalar.  
The factor of two is due to the two polarizations 
in the usual sum over modes.

For general $\gamma$ we use eq (\ref{rtg})and 
numerical integration gives the following:\vspace{2mm} \newline\begin{tabular}{|c||l|l|l|l|l|l|l|l|l|l|}\hline {$\gamma$} & {$0^o$} & {$5^o$} & {$10^o$} & {$15^o$} & {$20^o$} & {$25^o$} &{$30^o$}&{$35^o$}&{$40^o$}&{$45^o$}\\ \hline {$E/E_0$}&  1.000 & 0.984 &  0.951 & 0.907 & 0.856 & 0.800 & 0.741 & 0.683 & 0.624 & 0.569  
\\ \hline \end{tabular}  \begin{tabular}{|c||l|l|l|l|l|l|l|l|l|}\hline {$\gamma$} & $50^o$ & 
{$55^o$}&{$60^o$}&{$65^o$}&{$70^o$}&{$75^o$}&{$80^o$}&{$85^o$}&{$90^o$}\\ \hline {$E/E_0$}& 0.515& 0.468 & 0.425 & 0.387 & 0.356 & 0.330 & 0.313 & 0.301 & 0.2984897749 \\ \hline \end{tabular}\vspace{1.6mm}
 fig 1 represents the same $E(\gamma)/E_0$ data graphically
\begin{figure}[h]
\centerline{\psfig{figure=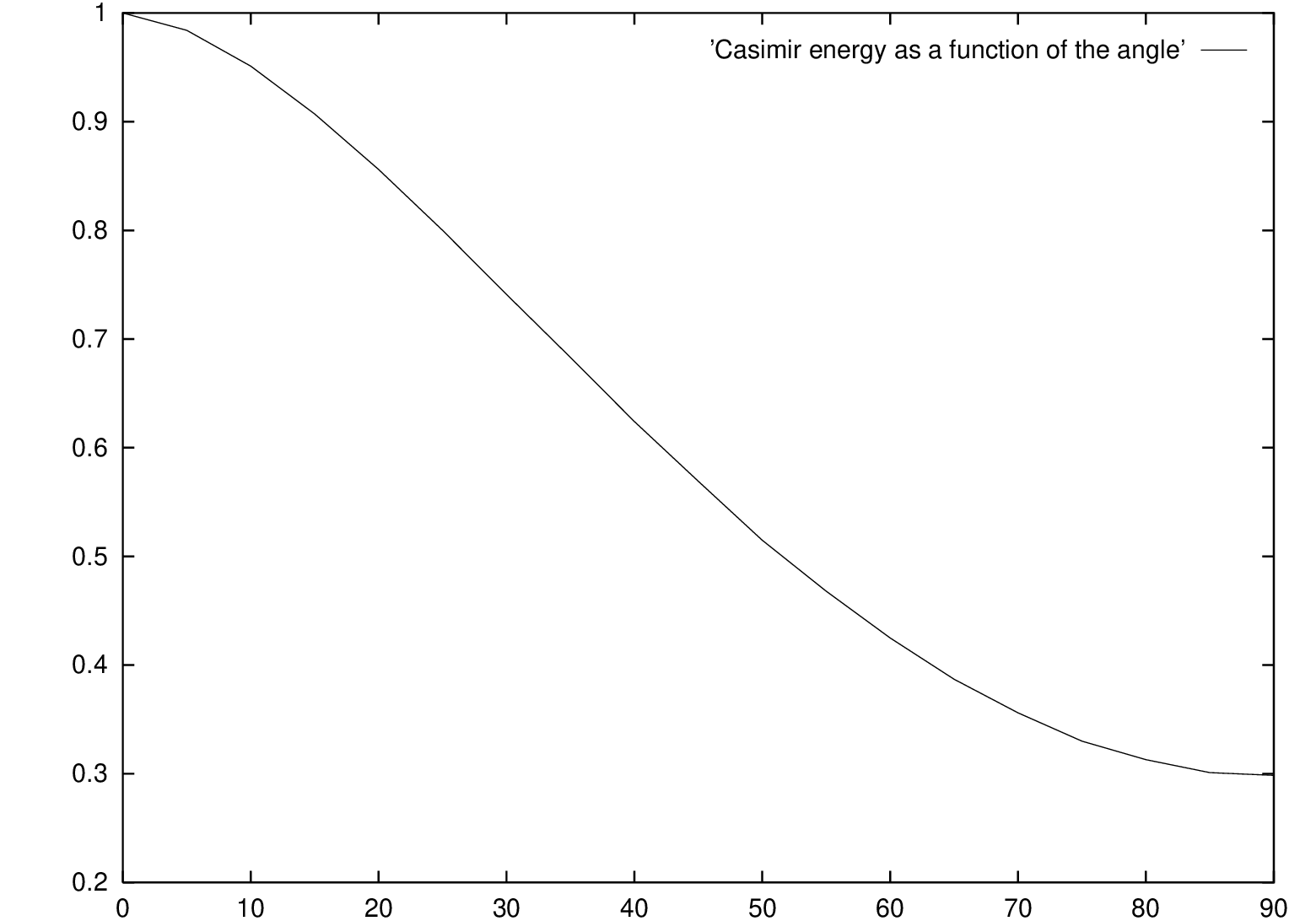,width=14cm,height=8cm,clip=}}
\label{fig1}
\end{figure}

Naturally occurring materials conducting only along some preferred 
directions would have provided an ideal
 setting for testing the angle dependent effect.  
Indeed such materials  would have manifested the unisotropic 
conductivity already on atomic
Angstrom scales. We are not aware of any adaquate candidates.
\footnote{ The materials
 exhibiting the high $T_c$ phenomenon have a layered structure and tend to
 conduct mainly in the planes of the layers.  If we would cut such materials
 by a plane perpendicular to these layers then we would obtain a surface with
 a striped pattern of conduction.  However, precisely because of the fact that
 one is not cutting along a natural cleavage direction the required smoothness
 of surfaces cannot be thus achieved. }
 Fortunately unisotropic conductivity
 at atomic scales is not required for verifying our effect. 
 The Casimir forces are intimately related to the Casimir 
Polder interactions.  This implies a similar (retardation) lower bound for the interplate distance
 $a$, at which the Casimir force will not be masked by other effects:
\beq
a \geq a_{min} \approx {d_{atom}\over (v_F/c)}
\eeq

where $d_{atom}$ is a relevant dimension, say the inter atomic
 distance 
in the crystal of order (few) Angstroms, and    $v_F$      
 is the Fermi velocity.
  For the conduction electron in gold the latter is of the order of          
  $10^8$ cm/sec, and hence  $a_{min} \approx 0.1 \ \mu$.
        Indeed measurements of the ordinary
 Casimir effect were performed at yet larger  distances:
   $0.5\mu<a<5\mu$     and  $0.2\mu<a<2\mu$ 
     in \cite{L} and \cite{M}  respectively.

        The interplate separation, $a$, sets the basic scale and cutoff
 for the Casimir problem.  
A fact, which as we will elaborate on next, is crucial for the
 proposed new experiments.  Our suggestion is to obtain the unisotropic
 conductivity via ``striped coating'' by a conducting layer.  
The above suggests that taking the distance between the stripes, $d$, 
to be say $d\simeq a/2$
                         will suffice for the purpose of testing the new 
angle  dependent effect.                                         
Indeed expanding equation (\ref{fr2}) above we have:
\beq
{E\over A}= \sum_n\int d^2\hat k {[\alpha(\hat k)]^{2n}\over n} \int_0^\infty
dk k^2 e^{-2nka}
\eeq
Integrating over $k$ from $0$ to $\infty$ yields
\beq
{E\over A} = \sum_{n=1}^\infty d^2\hat k [\alpha(\hat k)]^{2n} {1\over n^48a^3}
.\eeq
Both the $n$ sum {\em and } the $k$ integrations above are rapidly convergent. 
 Specifically, the $n = 1$  term contributes at least
 $90/\pi^4\approx 90\% $
  of the total $E/A$.  (This happens for   $\gamma=0$    and
$\alpha=1$.        
          In general   $\alpha<1$       and the relative contribution of the
$ n > 1$ terms is further suppressed ). 
The $k$                   integration is
 exponentially suppressed once  $k> a^{-1}$     .  
Since modes of wave number
$k$  can resolve surface details (or the induced currents $J_\mu(x)$        )
 only at scales  $\Delta x\sim k^{-1}$, 
the unisotropy of the conductivity needs
 to manifest only at a distance scale $\approx a$.    
This feature alongside the recently performed ordinary Casimir 
experiments motivates ours
 suggested experiments for testing the new angle dependent effect.
 
To achieve the required smoothness of surfaces and  sensitivity, and in order
 to avoid possible domination of the forces by random, nearby contact points,
 the recent (and older! \cite{SP}) Casimir experiments all share the following 
three elements:

1.    ``Perfect'' quartz surfaces are coated by thin
 layers (of thickness $\approx 0.1 \ \mu$              ) 
of conducting metals (say gold) and thus
 form the required smooth conducting surfaces

2. Rather then having two parallel plates the experiments utilize one 
flat plate and a plano-convex
 spherical lens.  The radius of curvature of the latter R 
is much larger than 
$a$  the distance at the point of nearest proximity \footnote{
 Under these circumstances one can evaluate the casimir energy force by
 integrating over concentric annuli.  This yields on effective plate area
$ A=\pi L^2$ with $ L=\sqrt{2Ra}$}

3. Like in all sensitive measurements of small 
forces, dating back to the Cavendish  measurement
 of $G_N$, a torque balance is used.

Our suggested methodology parallels very
 much the above.  The key new element is that the coating is done along parallel
 stripes in the plane of the plate, say along the $x$ axis, with distance $d$
 between the centers of the stripes.  More precisely we have a fraction $f$
 of the original quartz surface covered and a fraction ($1-f$) is left
 uncovered.  The width of the
 conducting stripes (insulating intervening stripes) are then               
  $fd$ (and $(1-f)d$ ),     respectively.  
Even if $d$ is $0.5 \ \mu$               and    $f=0.5$,                  
  present MBE (molecular beam epitaxy) and other (nano-) technologies enable
 us to generate the required uniform parallel stripes.
                   Also this striped coating can be done over areas 
comparable with the
``effective plate area'' contributing in the above Casimir
 measurements.The angle dependent force $F/A\sim W(\gamma)/a^4$
allows a more detailed verification of the Casimir
Phenomena.

To ensure its feasibility we have however to verify that the
``Standard obstacles'',familiar from the ordinary Casimir
experiments,and some new difficulties peculiar to the striped
variant,can be avoided in the new context. The rest of this note is
devoted to discussing these issues.
\newline{\bf 1)Finite conductivity corrections.}\newline
The $\omega$ dependence of $\epsilon$ and $\sigma$  implies that the latter tend to vanish for frequencies higher
 than the plasma frequency $\omega_p=(4\pi ne^2/m_e)^{1/2}$.
 Modes with $\omega >\omega_p$ are effectively not reflected and do
 not contribute to Casimir effects. Unless this intrinsic cutoff is
 larger than $1\over a$:
\beq\label{rvh} c/\omega_p={c\over (4\pi ne^2/m_e)^{1/2}}\leq a\eeq
 appreciable
 finite conductivity corrections \cite{SCH} and a reduction of the strength
 of the casimir effect cannot be avoided.  Since the density of conduction electrons, n,
 and the other parameters in $\omega_p$  cannot be varied for bulk materials, eq.
 (15) can be viewed as yet another lower bound, in addition to the
 "retardation bound" of eq(\ref{rvh}), on the plate separations for which
 the effect can be measured.  If we use  $n\simeq d_{atom}^{-3}$                  for the density
 of (gold) atoms and corresponding conduction electrons and also estimate
 via the virial theorem  $\langle m_ev^2\rangle\simeq
 m_ev_F^2\simeq\langle {e^2\over r}\rangle\simeq{e^2\over d_{at}}$,                                                  
            the limits (12) and (15) become the same, up to a numerical
 coefficient.  The last bound becomes more stringent when only a fraction f
 of the, say, quartz surface is covered with a gold layer of thickness $t$. 
 In this case an infalling electromagnetic  plane wave – which in bulk
 matter could have coherently interacted with all conduction electrons down
 to a depth of $l_{eff}\simeq 1/k\simeq c/\omega$
     , interacts only with a fraction
$ft/l_{eff}=ft\omega/c$     thereof.  The effective
 smeared electron density that should be substituted in this case for
 n in the expression for $\omega_p$, should therefore be accordingly
 reduced:$n\rightarrow n'=ntf\omega/c$ .  In this case even lower frequencies,
 exceeding $\omega_p'=\sqrt{4\pi n e^2ft\omega\over m_ec}$

or equivalently:
\beq \omega_{cutoff new}={4\pi ne^2ft\over m_ec}\eeq

will cease to contribute
 to the Casimir effect.  The discussion following eq. 12 of the connection
 between the separation $a$ and the relevant contributing modes            
 with $ k=\omega/c$  then implies the, new, finite conductivity corrected (F.N. 10),
 effective lower bound on the plate separation a:
\beq a\geq{\pi c\over\omega_{co new}}\simeq{1\over  4\lambda_e\alpha_{em}ntf}\eeq
                                                         (17)
with $
\alpha_{em} =1/137$ and $\lambda_e = \hbar/(m_ec)$    
                                       .
    The key point we
 would like to make is that also the last bound still allows our
 suggested measurements down to plate separations of $a= 0.5\mu$ 
  Thus for concreteness take the case of thickness $t=0.01\mu
 $ gold coating of half the surface.  In this case
    $ n={\rho_{AU}\over A_{AU}}N_{av}\simeq 6. 10^{22}.$ and                               and eq. 16 reads

\beq a/\mu\geq 0.3(f/({1\over 2}))(t/0.01\mu)\eeq

{\bf 2)  Possible Anderson localization effect due to the one
  dimensional stripes}\newline
At sufficiently low temperatures  One dimensional conductors exhibit  Anderson
 localization.  Since we may wish to perform the experiments at low
 temperatures one may wonder if this can effect the striped
 plate conductivity and hence the proposed experiment.
To show
 that these difficulties do not affect the proposed experiment we invoke
 again the discussion above of the relevant  $ J_\mu(x)$                configuration
 contributing to the casimir effect.  In terms of the Fourier
 transform variables these are $\tilde{J}_\mu(k)$ with $ k=1/a$.
   Hence we do not
 need the stripes to be conducting along the full length L of the
 macroscopic sample.  Rather it suffices that the stripes will be
 good conductors along distances $l$ of order several times $a$. 
 Since the distance between stripes $d$ and the width of individual
 stripes $ fd$ and thickness $t$      are also of order $a$
 (or rather smaller by a factor of 10 -100 ) the required
 conductivity of the stripes is practically guarenteed.

{\bf 3)Possible effects from changing overlap of stripes
 (or other pattern).}
A basic requirement in our proposed "set-up"
 is that the separation between the plates $a$ be larger than the
 separation/width of the stripes.  In this case the modes relevant for
 the Casimir effect of wavelength $ \approx 2a$            cannot resolve individual
 stripes.  Hence  $ W(Cas)$   will not depend in this case on the relative
 positioning of the stripes in the two plates but only on     
  the angle between their directions $\gamma$, which is exactly the effect of
 interest.  The decreasing (in absolute value) of  $ W(\gamma)$
 with $\gamma$      
   and the resulting tendency of the plates to align their stripes
 (or whatever other common pattern of conducting patches exist on them), 
 is a generic feature existing also when, $d$, the characteristic stripe
 (patch) size is larger than $a$.  The new angle $\gamma$      dependence follows from the
 fact that in this limit the total casimir energy is proportional
 to the total overlap area $A(overlapp)= gfA $ with $A=L^2$ the
 complete plate area, f the fraction of conducting stripes/patches on
 each conductor and g is the the'' overlap factor''.  If the patterns
 on both plates are identical (so that for $\gamma=0$               ,
 say, $ g=1$) the overlap factor decreases with increasing $\gamma$
 and so should $W_{Cas} (\gamma)$     
          for $d > a$.  It is amusing to note that the change of $W(\gamma)$  
 in this case is qualitatively different from that in the  $a > d$ the case of
 interest here, where as indicated in fig.         $ W(\gamma)$ varies
 smoothly as $\gamma$,the angle between the directions of conductivity
 changes from $0^o$ to $90^o$.\newline 
                    As one can readily verify the overlap of identical
 elements of size $d$ (and also stripes of seperation and width of
 order  $d$) varies dramatically: decreasing
 from $g(\gamma=0) = 1$ to its minimal value once $\gamma\approx d/L$                      .
  In passing we note that this dramatic sensitivity \footnote{This remarkable sensitivity
 to small rotations and also displacements of identical patterns is
 at the core of a  methods for verifying such patterns, e.g., in the case
 of finger-print recognition \cite{JJ}.}  could conceivably
 be utilized for yet another measurement of the ordinary Casimir effect. 
 The resulting torque in this case is ${\Delta W\over \Delta/gamma}= {\hbar
   c\pi^2L^3f\over 720a^4}$ and is compareable to the torques manifesting in the torqe ballance meassurements of the ordinary Casimir effect.\newline

Summing up we see then no obvious
 barrier to the proposed experiment.  The feature of having angular 
      dependence rather than only $a$ dependent forces is rather welcome.
      Even if we use very thin coating stripes
 (of width $t=0.01\mu$  which may be technologically advantageouss, as
 no high, unstable grooves are required, the finite
 conductivity corrections only mildly influence the angular dependence
 of $F_{Cas} (\gamma,a)$.
 In this case the first n = 1 term in (12) will dominate
 the sum for E/A even more strongly since higher k values are cut off.  While 
the force itself may be suppressed in proportion to $t/t_{crit}$                 with $t_{crit}$               
 given by saturation of the bound in eq.( 17), the angular dependence will be
 completely predictable and measurable.

\noindent
Acknowlejment

We enjoyed discussions with E. Ben Jacob, G. Deutcher, J. Feinberg, R. Jaffe,
E. Kapon, A. Mann, R. Onofrio, M. Revzen, and A. Palevski.

\end{document}